\documentclass[entropy,article,accept,moreauthors,10pt,a4paper]{mdpi} 
\firstpage{1} 
\articlenumber{x}
\doinum{10.3390/------}
\pubvolume{xx}
\pubyear{2017}
\copyrightyear{2017}
\history{Received: 29 May 2017; Accepted: 12 July 2017; Published: 14 July 2017}

\usepackage{color}
\usepackage{amssymb}
\usepackage[labelformat=simple]{subcaption}

\DeclareCaptionLabelFormat{subcaptionlabel}{\normalfont(\textbf{#2}\normalfont)}
\usepackage{lscape}
\graphicspath{ {Figures/}} 
\Title{Cosmological Time, Entropy and Infinity}

\Author{Clémentine Hauret *$^{,\dagger}$, Pierre Magain  $^{\dagger}$ and Judith Biernaux}

\AuthorNames{Clémentine Hauret, Pierre Magain and Judith Biernaux}

\address[1]{%
STAR-OrCA (Space sciences, Technologies and Astrophysics Research - Origins in Cosmology and Astrophysics), Universit\'e de Li\`ege, All\'ee du 6 Ao\^ut 19c, B-4000~Li\`ege, Belgium; pierre.magain@ulg.ac.be (P.M.);  jbiernaux@ulg.ac.be (J.B.)}
\corres{Correspondence: clementine.hauret@ulg.ac.be; Tel.: +32-4-366-9678}

\firstnote{These authors contributed equally to this work.} 
\abstract{Time is a parameter playing a central role in our most fundamental modelling of natural laws. Relativity theory shows that the comparison of times measured by different clocks depends on their relative motion and on the strength of the gravitational field in which they are embedded. In standard cosmology, the time parameter is the one measured by fundamental clocks (i.e., clocks at rest with respect to the expanding space).~This proper time is assumed to flow at a constant rate throughout the whole history of the universe. We make the alternative hypothesis that the rate at which the cosmological time flows depends on the dynamical state of the universe. In thermodynamics, the arrow of time is strongly related to the second law, which states that the entropy of an isolated system will always increase with time or, at best, stay constant. Hence, we assume that the time measured by fundamental clocks is proportional to the entropy of the region of the universe that is causally connected to them. Under that simple assumption, we find it possible to build toy cosmological models that present an acceleration of their expansion without any need for dark energy while being spatially closed and finite, avoiding the need to deal with infinite values.}

\keyword{cosmology: theory; entropy; arrow of time; dark energy}

\begin{document}

%%%%%%%%%%%%%%%%%%%%%%%%%%%%%%%%%%%%%%%%%%
%% Only for the journal Gels: Please place the Experimental Section after the Conclusions

%%%%%%%%%%%%%%%%%%%%%%%%%%%%%%%%%%%%%%%%%%
\section{Introduction}

Since its introduction nearly a century ago \citep{einstein1916}, general relativity (GR) has been brilliantly confirmed by a number of observations, most notably the perihelion precession of Mercury, the gravitational redshift, and the deflection of light by massive bodies.~GR has also been used in cosmology to describe the evolution of the universe as a whole, and the model that currently gives the best description of its large-scale structure and evolution (namely, the lambda cold dark matter ($\Lambda$CDM) model) is based on the equations of GR. However, in order to provide an accurate description of large-scale (galaxy-scale and higher) and long-term phenomena, a number of ingredients had to be added to the theory. The motion of stars in the outskirts of spiral galaxies and the velocity dispersions of galaxies in clusters required the addition of dark matter. The acceleration of the expansion of the universe called for the introduction of a repulsive component called dark energy.

One may note that all the successful tests of GR are dealing with small scales and are quasi-instantaneous in terms of cosmological time. Conversely, additional ingredients (dark matter and dark energy) are necessary to bring the models into agreement with observations dealing with very large scales and very long-term phenomena. We are thus faced with the following alternative: either these dark components are actual constituents of the universe (in which case they should sooner or later be identified) or there is something wrong with the application of GR on very large scales, and especially on the universe as a whole.

The vast majority of specialists are in favour of the existence of dark matter and dark energy. However, despite considerable effort, these dark components have not yet been identified. Some alternative theories have been proposed, the most popular being the modified Newtonian dynamics (MOND),~\cite{milgrom1983}.

In this paper, we want to explore another hypothesis---that the problems encountered when dealing with large scales are related to our fundamental understanding of time. Special relativity has shown that the time measured by different clocks depends on their relative motion.  GR has shown that the time they measure also depends on the strength of the gravitational field in which they are embedded.  We wish to go one step further and postulate that time depends on the dynamical state of the universe in which it is set.

The arrow of time is often put in relation with the second law of thermodynamics, which states that the entropy of an isolated system can only increase (or at best stay constant) as time flows.  In this context, we  propose that time might be regarded as an emergent property in the universe, proportional to the entropy of the causally connected volume, considered as the {\em ultimate isolated system}. Note that other theories dealing with emergent time \cite{Page1994} have been presented in the past. More recently, emergent gravity \cite{Verlinde2016} was also introduced in the context of string theories.

Starting from this assumption, we develop a very simple, semi-empirical, and mathematically basic toy model of our universe. Our goal in this paper is to design a straightforward implementation of our hypothesis and to determine its potential consequences on the (dark matter and dark energy) content and evolution of our universe. We thus emphasize that this study is only preliminary. However, as the results are promising  with an acceleration of the expansion of the universe without recourse to dark energy, this model is worth exploring further in forthcoming papers---particularly since other theories investigating the link between entropy and the accelerated expansion of our universe have recently been~proposed \cite{Pavon2014}\cite{Pandey2017}%\cite{Pavon2014,Pandey2017}.

For the purpose of this paper, we must make a distinction between two concepts of time: the physical one (also called cosmological time in this paper) and the coordinate one. We will define and characterise them in further detail in Section  \ref{Sec-Times}. After that, in Section \ref{Sec-Ent_universe}, we will specify our assumption on the relation between the dynamical state of the universe (here, the entropy) and the (cosmological) time concept. We stress that we only present in this article one example among others of practical implementation of this link. Obviously, other mathematical dependences between time and entropy or between time and a parameter characterising the dynamical state of the universe (such as its curvature or its density parameters, for example) might be proposed or---even better---derived from theoretical considerations. Nevertheless, in Section  \ref{Sec-Evol_cosmo_time} we build a semi-empirical toy model and show that it can produce a spatially closed universe with an accelerating expansion without dark energy.

\section{Cosmological Time and Coordinate Time}
\label{Sec-Times}

When applying GR to the whole universe in order to derive the $\Lambda$CDM model, one assumes that it is homogeneous on large scales and that the small-scale inhomogeneities have no impact on its evolution as a whole.  One also assumes that the four-dimensional space-time can be sliced into three-dimensional space-like hypersurfaces labeled with a time parameter $t$.  This time parameter is usually taken as the time measured by fundamental comoving clocks \cite{hobson2006}.  The time $t$ is then identified with the proper time of fundamental observers.  It is generally taken for granted that if the universe is homogeneous, this proper time flows at the same rate all along the history of the universe. When Einstein derived the basic theory of GR, the universe was believed to be essentially static and there was no reason to question that hypothesis.  However, in the framework of an evolving universe, one might ask whether the rate at which time flows could depend on its dynamical state.  

We thus propose to distinguish between two different times: (1) the (conventional) coordinate time parameter $t$, which is the one measured by fundamental clocks at the Big Bang and is assumed to flow at a constant rate along the evolution of our universe; and (2) the cosmological time $\tau$, which is the one measured by fundamental clocks at any times and is assumed to depend on the state of the universe and to control all physical processes. 

\begin{figure}[H]
\centering
\includegraphics[width=0.6\linewidth]{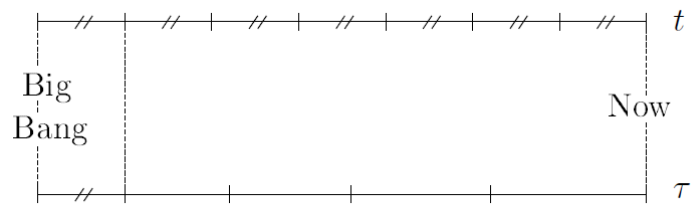}
\caption{\label{fig-sch-times} Illustration of the difference between the cosmological time $\tau$, of which the units are of varying length, and the coordinate time $t$, which flows at a constant rate. The unit of $t$ is chosen to be equal to the unit of $\tau$ at the Big Bang.}
\end{figure}

To sum up, this $\tau$ time is the time we live in. Every measurement is thus made in this time. However, the unit of $\tau$ varies with the evolution of the universe (i.e., a unit of $\tau$ at time $\tau_1$ is not equal to a unit of $\tau$ at time $\tau_2 \ne \tau_1$), as can be seen at the bottom of Figure \ref{fig-sch-times}. On the other hand, the coordinate time $t$ is chosen so that its unit does not vary with time. To set the unit of $t$, we decide to make it equal to the unit of $\tau$  at the Big Bang, as shown in this same Figure \ref{fig-sch-times}. 

Cosmological models with varying fundamental constants such as the gravitational constant $G$, the fine-structure constant $\alpha$, or the speed of light $c$ have already existed for a long time. In 1957, Dicke  published one of the first theories of gravitation with variable constants \cite{Dicke1957} and a review of the numerous models developed since then can be found in \cite{Chiba2011}.  In this paper, we go one step further with a varying time flow.

\section{Cosmological Time, Ultimate Isolated System, and Entropy of the Universe}
\label{Sec-Ent_universe}

In the following and as already stated, we make a specific hypothesis on the relation between the cosmological time $\tau$ and the physical properties of the universe to develop a mathematically rudimentary cosmological model. 

It is often argued that the direction of time flow---the arrow of time---is dictated by the second law of thermodynamics: the direction of time flow is the one for which the entropy of an isolated system increases.  Let us consider fundamental observers in the universe.  The largest system possibly interacting with such observers (i.e., the ultimate isolated system from their point of view) is the region of the universe that is causally connected to them.  That region is bounded by the particle horizon, defined by the distance a light signal could travel from the beginning of time to the observers.  From their very point of view, that system can be considered isolated since no interaction can happen with objects located further than the particle horizon.  We  thus make the simple assumption that the cosmological time $\tau$ measured by such observers is proportional to the entropy of the region of the universe that is causally connected to them.

Conventional estimates of the entropy budget (e.g.,\  \cite{egan2010}) result in black holes dominating the entropy of the universe by far.  Indeed, if one admits the Bekenstein--Hawking estimate for the black hole entropy \cite{bekenstein1973}\cite{hawking1976}%\cite{bekenstein1973,hawking1976}
, the entropy budget of the stellar black holes would exceed that of the (much more numerous) stars by 17 orders of magnitude and that of the cosmic microwave background (CMB) by seven to eight orders of magnitude \cite{egan2010}.  Even more dramatically, a handful of supermassive black holes like those found at the centres of galaxies would completely dominate the entropy budget of the~universe.

We do not consider these possible sources of entropy in our calculations for the following reason.  Let us consider the typical situation in which a stellar black hole can be detected.  This is a binary stellar system in which one of the stars is a red giant losing matter to its compact companion.  An accretion disk forms around the compact object and emits high energy radiation as its temperature reaches high values close to the central object.  Whether the compact object is a neutron star or a black hole can only be known by estimating its mass and verifying whether or not it exceeds the maximum mass of neutron stars ($\sim 3 M_{\odot}$). The gravitational effect of a black hole just above that mass limit on the accretion disk (and even more, on the causally connected region) hardly differs from that of a neutron star just below the mass limit. However, in the Bekenstein--Hawking framework, there would be a dramatic entropy discrepancy between those two objects, without any measurable effect on the causally connected region, let alone on the cosmological time. To avoid that inconsistency, we choose not to consider the Bekenstein--Hawking entropy. Even if we did, it is reasonable to assume that the black hole entropy has no effect on the entropy budget outside of its horizon. Regardless, this hypothesis could always be rewritten such that cosmological time is proportional to the radiation entropy in the causally connected region of the universe. For any of these three reasons, we chose not to consider black holes' entropy in our calculation.

The entropy of the universe is hence dominated by the CMB \cite{egan2010}.  Relic neutrinos are predicted to contribute nearly as much as the CMB.  However, as their entropy has the same dependence on temperature and volume as the CMB photons, including them would only change the proportionality coefficient between entropy and cosmological time, and not the relation between cosmological and coordinate times.

\section{Evolution of Cosmological Time}
\label{Sec-Evol_cosmo_time}

The CMB is a photon gas very close to thermodynamic equilibrium, whose entropy $S$ only depends on its volume $V$ and temperature $T$ via \cite{egan2010}
\begin{equation}
\label{eq-S}
S = \frac{4 \pi^2 k_B^4}{45 c^3 \hbar^3} V T^3,
\end{equation}
where $k_B$ is the Boltzmann constant, $c$ the speed of light and $\hbar$ the reduced Planck constant. $V_{\rm horiz}$ being the causally connected volume, we assume $\tau \propto S(V_{\rm horiz}) \propto V_{\rm horiz} T^3$. To compute the causally connected volume, we solve the Einstein equations of GR in a metric representing our universe.

By construction, GR is a local theory.  Applying it to the whole universe is an attempt to use a local theory to solve a global problem.  If the universe is spatially homogeneous, applying the Einstein equations in the vicinity of any observer is equivalent to applying them anywhere else, which means this local theory may be extended to any point in space.  However, in our toy model, this is not true for any instant in time, as the cosmological time $\tau$ will not flow at a constant rate with the evolution of our~universe.

As GR has been successfully verified locally, we make the hypothesis that the Einstein equations are valid at any location in space-time, when written as a function of the local space-time coordinates $(t, r, \theta, \phi)$.  Then, under the assumptions of homogeneity and isotropy of the universe, the interval $ds$ is given by the Robertson--Walker metric:
\begin{equation*}
   ds^2 = c^2 dt^2 - R^2 \left[ \frac{dr^2}{1-kr^2} + r^2 (d\theta^2 + \sin^2\theta \, d\phi^2 ) \right],
\end{equation*}
where $( r, \theta, \phi)$ are the comoving spherical coordinates, $R$ is the scale factor of our universe, and $k$ measures its spatial curvature.  The Robertson--Walker metric and the Einstein equations are thus assumed to be valid at any point in space and at any time.~However, the proper time flow of fundamental comoving observers will vary with time.  After solving the equations for any local proper time $t$, we must connect these local proper times together by replacing them with the cosmological proper time $\tau(t)$.  The equations cannot be directly solved as a function of $\tau$, as the latter depends on the global properties of the universe and is not an independent variable.  Of course, and as already pointed out, such a procedure should later be tested in a more rigorous mathematical framework, as the aim of the present paper is only to illustrate the consequences our basic hypothesis might have on the content and evolution of the universe.

Hence, in this specific implementation of our basic assumption, we have to solve the Einstein equations in coordinate time $t$, as it is the only time we can access at this stage of the calculations. Through this, we aim to obtain the evolution (still in $t$ time) of the entropy in the causally connected volume $S_{\rm horiz}(t)$ to apply our assumed relation between the cosmological time $\tau$ and this entropy ($\tau(t) \propto S_{\rm horiz}(t)$). Thanks to this assumption, we can then associate each coordinate time $t$ (and thus each value of the scale factor $R$) with the cosmological time $\tau$, which effectively controls the physical phenomena. The resolution of equations in coordinate time $t$ (i.e., a resolution of the usual Friedmann equation) is thus only (in the toy model presented here) a way to determine the physical time $\tau$.

The Einstein equations in $t$ are simplified to the usual Friedmann equation. The analytical solutions $t(R)$ can be easily found in the case of a universe where matter dominates the matter--energy budget (as is the case now) and used to determine the comoving distance of the horizon $r_\mathrm{horiz}$:
\begin{equation}
\label{eq-r_horiz}
r_\mathrm{horiz} = \frac{1}{\sqrt{|k|}} ~\Theta_k \left( \sqrt{|k|} \int_0^{R} \frac{c}{R} \frac{dt}{dR} dR \right)
\end{equation}
where the $\Theta_k(x)$ function is defined according to the sign of $k$:
\begin{equation*}
\begin{array}{cc}
& \\
\Theta_k(x) & =  \\
& 
\end{array} 
\left\{
\begin{array}{cccc}
 \sin (x) & \mathrm{if~} k > 0 \\
 x & \mathrm{if~} k = 0 \\
 \sinh (x) & \mathrm{if~} k < 0
\end{array}
\right. .
\end{equation*}
We can solve Equation (\ref{eq-r_horiz}) to obtain a simpler expression of $r_\mathrm{horiz}$ as a function of the present Hubble constant $H_{0, t}$ and of the present matter-density parameter $\Omega_{0,t}$, both defined in coordinate time $t$ by
\begin{equation}
\label{eq-Hub_dens_t}
H_{0, t} = \frac{1}{R_0} \frac{dR}{dt} \Big|_{t = t_0} ~~ \mathrm{and}~~\Omega_{0,t}  = \frac{8 \pi G \rho_0}{3 H_{0, t}^2}
\end{equation}
where $G$ is the gravitational constant and $\rho_0$ the present matter density. We also set the present scale factor value $R_0$ to unity. $r_\mathrm{horiz}$ is then given---regardless of the spatial curvature of the universe---by
\begin{equation*}
r_\mathrm{horiz} = \frac{2c}{H_{0,\mathrm{t}} \Omega_{0,t}} \sqrt{R} \sqrt{\Omega_{0,t} + R (1-\Omega_{0,t})}.
\end{equation*}

From that equation, we can compute the expressions of the causally connected volume $V_{\rm horiz} = 4 \pi R^3 \int_0^{r_\mathrm{horiz}}x^2 / \sqrt{1-kx^2} dx$, of the entropy $S_{\rm horiz}$ in this volume thanks to Equation (\ref{eq-S}), and of the temporal variation of this entropy $dS_{\rm horiz}/dt$ :
\begin{equation*}
\frac{dS_{\rm horiz}}{dt} = \frac{64 \pi^3 k_\mathrm{B}^4 T_0^3}{45 \hbar^3} \frac{1}{H_{0, t}^2 \Omega_{0,t}^2} \left( \Omega_{0,t} + R (1-\Omega_{0,t}) \right),
\end{equation*}
where $T_0$ is the present temperature of the CMB and where we have used the relation $TR=T_0R_0$. From that, we easily determine the relation between the two times $t$ and $\tau$, as we assume that $\tau \propto S_{\rm horiz}$ (see Section \ref{Sec-Ent_universe} and the beginning of Section \ref{Sec-Evol_cosmo_time}):
\begin{equation}
\label{eq-dtaudt}
\frac{d\tau}{dt} = \frac{dS_\mathrm{horiz}/dt}{dS_\mathrm{horiz}/dt|_{\mathrm{BB}}} = 1 + R ( \frac{1}{\Omega_{0,t}} - 1 ).
\end{equation}
The constant of proportionality  $ dS_\mathrm{horiz}/dt|_{\mathrm{BB}} $ stands for the temporal variation of the entropy in the causally connected volume at the Big Bang, so that the coordinate time and the cosmological time are set equal at the Big Bang. 

One can directly notice that $d\tau/dt = 1$ for a flat universe ($\Omega_{0,t} = 1$). Thus, in the geometrically flat case, the cosmological time flows at a constant rate along the evolution of the universe, $\tau$ and $t$ become identical, and our usual representation of time does not need to be modified. In non-flat cases, however, the distinction between these times must obviously be maintained. Therefore, as $\tau$ is the physical time we live in, we do not measure the present Hubble and matter-density parameters in $t$ time, but in $\tau$ time. They are defined as in Equation (\ref{eq-Hub_dens_t}), but as functions of the cosmological time:

\begin{equation*}
\label{eq-Hub_dens_tau}
H_{0, \tau} = \frac{1}{R_0} \frac{dR}{d\tau} \Big|_{\tau = \tau_0} ~~ \mathrm{and}~~\Omega_{0,\tau}  = \frac{8 \pi G \rho_0}{3 H_{0, \tau}^2}.
\end{equation*}
Combining these definitions in our two different times, we obtain:
\begin{equation*}
\Omega_{0,\tau}  = \frac{8 \pi G \rho_0}{3  \left( \frac{1}{R_0} \frac{dR}{d\tau} \Big|_{\tau = \tau_0} \right)^2} = \frac{8 \pi G \rho_0}{3  \left( \frac{1}{R_0} \frac{dR}{dt} \Big|_{t = t_0} \frac{dt}{d\tau} \Big|_{\tau = \tau_0} \right)^2} = \Omega_{0,t} \left( \frac{d\tau}{dt} \Big|_{t = t_0} \right)^2.
\end{equation*}
Using Equation (\ref{eq-dtaudt}) at present time (i.e., with $R_0 = 1$), we  compute the observed matter-density parameter $\Omega_{0,\tau}$ from the $t$ matter-density parameter $\Omega_{0, t}$:
\begin{equation*}
\label{eq-Rel_dens}
\Omega_{0,\tau} = \frac{1}{\Omega_{0, t}}. 
\end{equation*}

\begin{figure}[H]
\centering
\begin{subfigure}[t]{0.65\textwidth}
\centering
\includegraphics[width=\linewidth]{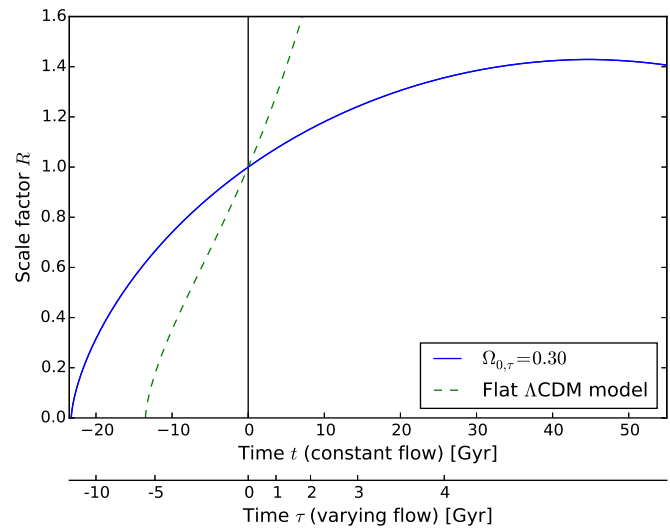}
\caption{\label{fig-acc-t-0.3}}
\end{subfigure}
\begin{subfigure}[t]{0.65\textwidth}
\centering
\includegraphics[width=\linewidth]{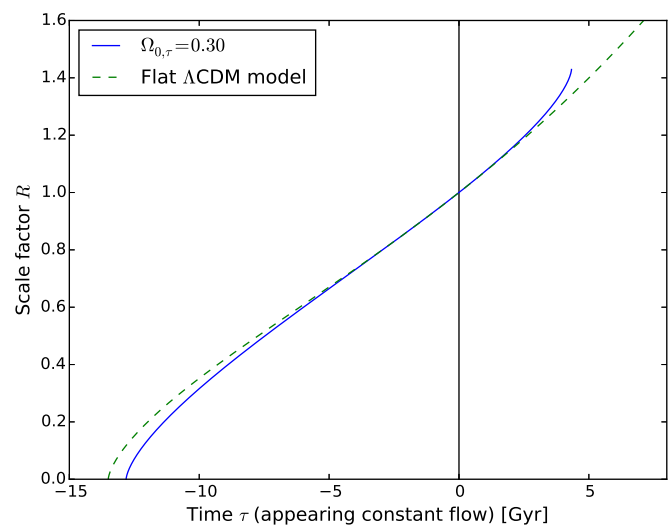}
\caption{\label{fig-acc-tau-0.3}}
\end{subfigure}
\caption{\label{fig-acc-0.3}Evolution of the scale factor $R$ of a universe as a function of the times $t$ and/or $\tau$ for our model (solid blue line; we choose to take $\Omega_{0,\tau} = 0.30$) and for a flat lambda cold dark matter ($\Lambda$CDM) model (dashed green line; we fix its matter density parameter at $\Omega_{0} = 0.30$ to be consistent with the most recent observations of type Ia supernovae \cite{betoule2014} or of the Planck telescope \cite{planck2015}). (\textbf{a})~When depicted as a function of the constantly flowing coordinate time $t$ or as a function of the varying cosmological time $\tau$, our model shows the usual behaviour of a spatially closed universe;  (\textbf{b})~When $\tau$ is wrongly interpreted as a constantly flowing time, this universe seems to accelerate. One will notice that the expansion of the universe reaches a maximum value of $R \approx 1.4$, corresponding to the maximum value of the scale factor of a closed universe (i.e., the maximum of the blue curve in Figure~2a). It does not continue to expand indefinitely, as in the case of a $\Lambda$CDM model. With $\Omega_{0,\tau} = 0.30$ and a Hubble constant fixed at $H_{0, \tau} = 70$ km/s/Mpc, our model predicts an age of this universe (as measured in cosmological proper time units) of $12.81$ billion years.} 
\end{figure}

\begin{figure}[H]
\centering
\begin{subfigure}[t]{0.7\textwidth}
\centering
\includegraphics[width=\linewidth]{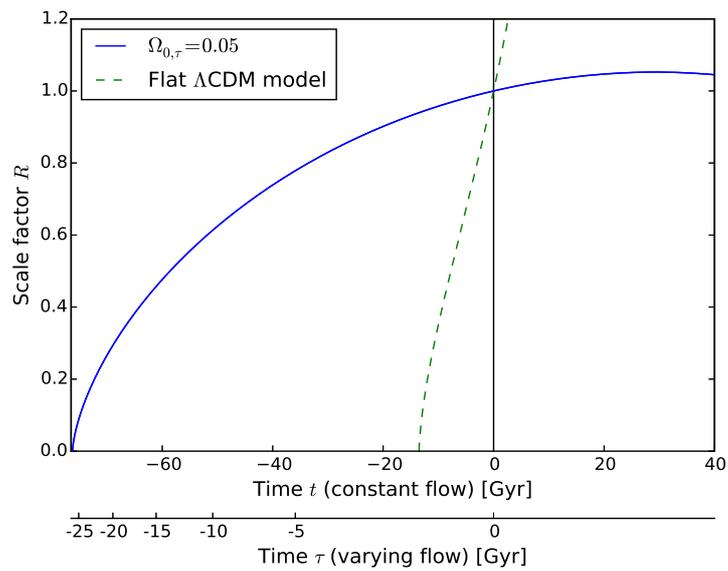}
\caption{\label{fig-acc-t-0.05}}
\end{subfigure}
\begin{subfigure}[t]{0.7\textwidth}
\centering
\includegraphics[width=\linewidth]{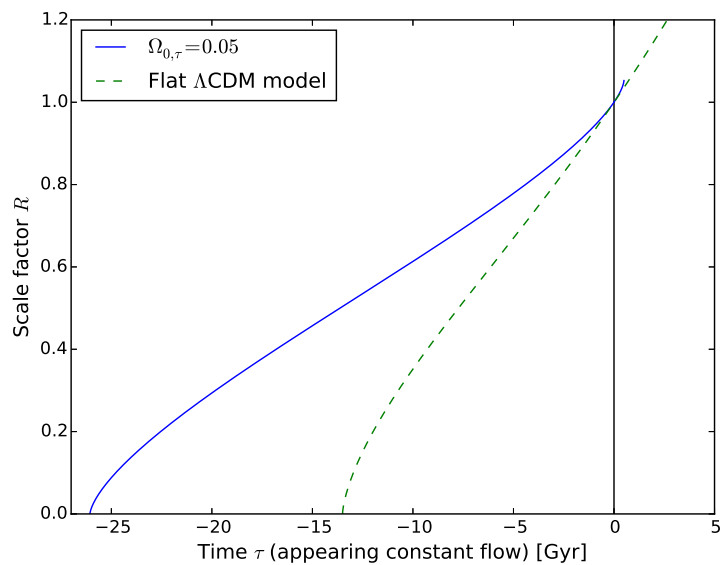}
\caption{\label{fig-acc-tau-0.05}}
\end{subfigure}
\caption{\label{fig-acc-0.05}Same as in Figures \ref{fig-acc-t-0.3} and \ref{fig-acc-tau-0.3} for a model with $\Omega_{0,\tau} = 0.05$. With this value of $\Omega_{0,\tau}$ and a Hubble constant fixed at $H_{0, \tau} = 70$ km/s/Mpc, our model predicts an age of this universe (as measured in cosmological proper time units) of $26.06$ billion years. The flat $\Lambda$CDM model has the same matter density parameter as previously.}
\end{figure}

A particularly interesting case concerns models with $\Omega_{0,t} > 1$, i.e.\  spatially closed cosmological models.  Indeed, the spatial curvature $k$ is given by $k = H_{0, t}^2 (\Omega_{0, t} -1)/c^2$ with negative/null/positive values of $k$ indicating spatially open/flat/closed universes. We illustrate this interesting case in Figures \ref{fig-acc-0.3} and \ref{fig-acc-0.05} where we plot the evolution of the scale factor $R$ of our toy universe as a function of our times for two particular values of $\Omega_{0,t} > 1$, hence $\Omega_{0,\tau} < 1$.

\begin{figure}[H]
\centering
 \includegraphics[width=0.8\textwidth]{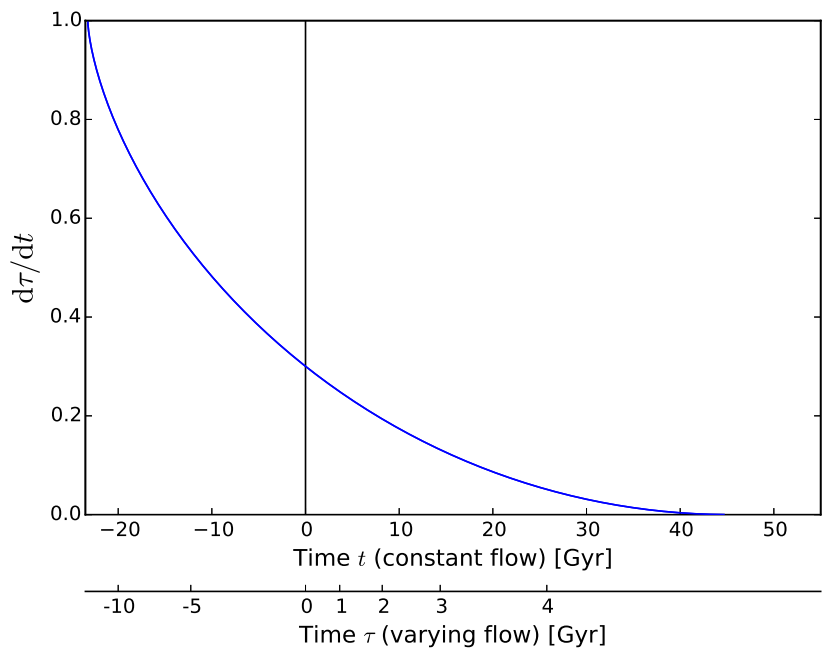}
 \caption{\label{fig-Evol_dtaudt-0.3}Variation of the cosmological time flow $d\tau /dt$ as a function of times $t$ and $\tau$ for a matter-dominated universe with $\Omega_{0,\tau} = 0.30$. The cosmological time $\tau$ continuously slows down in the present matter-dominated era.  The units of $\tau$ and $t$ are set equal at the beginning of the matter era when the universe was essentially flat, and $\tau = 0$ corresponds to the present epoch.}
\end{figure}

\begin{figure}[H]
\centering
 \includegraphics[width=0.8\textwidth]{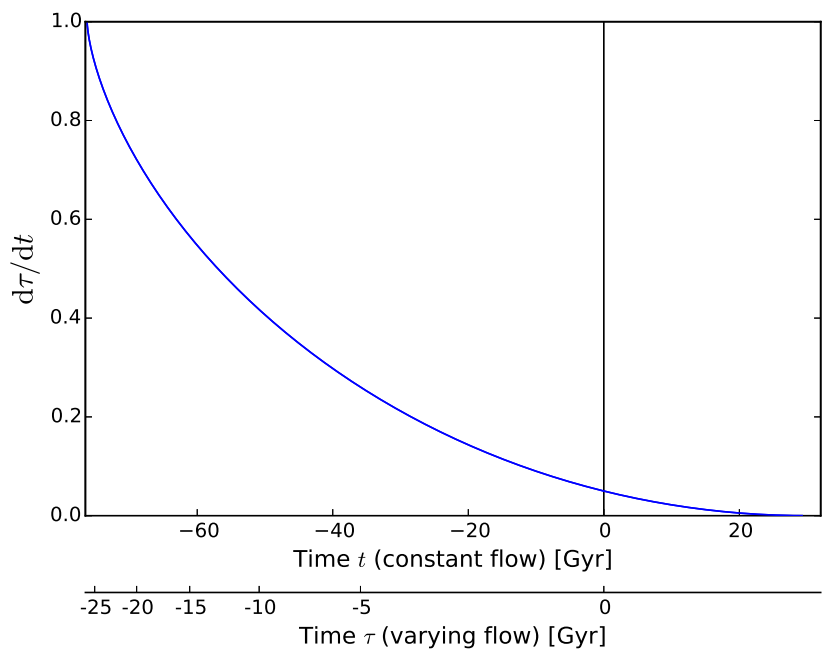}
 \caption{\label{fig-Evol_dtaudt-0.05} Same as in Figure \ref{fig-Evol_dtaudt-0.3} for a model with $\Omega_{0,\tau} = 0.05$.}
\end{figure}

In Figures \ref{fig-acc-t-0.3} and \ref{fig-acc-t-0.05}, this evolution as a function of $t$ is displayed  and one can directly notice the typical behaviour of closed models (an expansion phase with a maximum value of $R$, followed by a contraction phase towards a Big Crunch). In these same figures is also shown the correspondence (calculated via Equation \ref{eq-dtaudt}) between the constantly flowing time $t$ and its associated true varying time $\tau$ (abscissa axes of Figures \ref{fig-acc-t-0.3} and \ref{fig-acc-t-0.05}). Indeed, one can notice that a universe like the one we are currently dealing with (that is to say, with $\Omega_{0,t} > 1$) will have its $\tau(t)$ continuously slowing down (as displayed in Figures \ref{fig-Evol_dtaudt-0.3} and \ref{fig-Evol_dtaudt-0.05}),  while the opposite behaviour will obviously be observed in a universe with $\Omega_{0,t} < 1$. 

If the $\tau$ time is wrongly considered as having a constant flow (as it is the case if we try to represent an universe with the usual cosmological models), the $R(\tau)$ curve presents an apparent acceleration, as is shown in Figures \ref{fig-acc-tau-0.3} and \ref{fig-acc-tau-0.05}. This is a similar phenomenon to the transformation of a logarithmic axis of a chart to a linear one, a straight line in one representation of this axis being an exponential (accelerating) curve in the other one.

\newpage

Consequently, the main point of this article is that in our varying vision of time with the hypothesis that sets cosmological time proportional to the entropy of the causally connected region, we can produce a universe that seems to accelerate while being in reality spatially closed. This is an interesting result, as in usual Friedmann--Lemaître cosmological models, a universe with such an accelerating behaviour is interpreted as open and thus infinite. Hence, in our interpretation of time, a universe appearing in acceleration can in fact be closed and finite, saving ourselves from philosophical problems related to an infinite universe (i.e., containing an infinite amount of matter). 

Indeed, there is no question about the usefulness of the concept of infinity in mathematics. Positive integer numbers can be summed up without any limit, reaching towards infinitely large numbers.  Conversely, any positive real number can be divided by two, for example, without any limit, heading towards infinitely small numbers. However, this is not the case in the physical world.  Consider some macroscopic quantity of water.  You can always divide it in half, but you cannot do so an arbitrarily large number of times, because at some point you reach the atomic (or nuclear, or quark) level and you cannot divide any more.

The only object in the physical world that is {\em postulated} to be spatially infinite is the universe. Such a universe with---on average---a uniform density of matter, would mean an infinite number of atoms, stars, galaxies, etc.  It would imply that this infinite amount of matter would have appeared during the Big Bang. In our opinion, this makes little physical sense. Indeed, in a rigorous mathematical approach, infinity is immeasurable. A variable can tend towards it, but anything measurable, like a quantity of matter, cannot be equal to it. Yet, within the scope of our toy model, the apparent openness and infinity of the universe is simply an artefact emerging from the variable time flow, whereas its real nature corresponds to a geometrically closed universe.

\section{Conclusions}

As already stated in the introduction, our goal with this paper is only to design a simple implementation of our hypothesis  of variable time in order to determine its potential consequences on the content and evolution of the universe. 

From the basic assumption that the proper time of fundamental observers is proportional to the entropy of the region of the universe causally connected to them, we obtain cosmological toy models that present accelerating expansion without recourse to dark energy. They also display positive geometric curvatures, indicating finite and spatially closed universes.  We thus avoid the philosophical problem of an infinite universe, containing an infinite amount of matter.

We nevertheless want to stress that---as pointed out earlier---other mathematical dependences between time and entropy or any other parameter characterising the dynamical state of the universe might be proposed or even better derived.

Obviously, the next step for us is now to compare our model predictions to the various cosmological observations available. However, to put these tests into practice, we have to develop a complete and mathematically thorough cosmology. Our aim is to address this in forthcoming papers. 

With a correct metric in $\tau$, we will be able to calculate distances and in particular luminosity distances in order to test our hypothesis on the observed Hubble diagram built with standard candles such as type Ia supernovae. A priori, since our model shows an accelerating evolution, we can expect a similar behaviour as the one from a usual Friedmann--Lemaître model with dark energy. Moreover, the effects of the slowing down of proper time on the rotation curves of spiral galaxies and on the velocity dispersions of galaxies in clusters have to be investigated.  Qualitatively, one would predict an apparent acceleration---similar to the one observed for the scale factor $R$---in these systems, which might explain the larger than expected observed velocities. Of course, very careful calculations will be necessary to confirm this tendency. Finally, cosmological nucleosynthesis and CMB anisotropies will provide further tests of such a model.

%%%%%%%%%%%%%%%%%%%%%%%%%%%%%%%%%%%%%%%%%%
\newpage

%%%%%%%%%%%%%%%%%%%%%%%%%%%%%%%%%%%%%%%%%%
%% optional
%\supplementary{The following are available online at www.mdpi.com/link, Figure S1: title, Table S1: title, Video S1: title.}

%%%%%%%%%%%%%%%%%%%%%%%%%%%%%%%%%%%%%%%%%%
\acknowledgments{Judith Biernaux  acknowledges the support of the F.R.I.A.\  fund (Fonds pour la formation \`a la Recherche dans l'Industrie et dans l'Agriculture) of the F.R.S.\   - F.N.R.S.\  (Fonds de la Recherche Scientifique).
%Please use full names.
}

%%%%%%%%%%%%%%%%%%%%%%%%%%%%%%%%%%%%%%%%%%
\authorcontributions{Clémentine Hauret and Pierre Magain contributed equally to this work. Judith Biernaux contributed in discussions and manuscript improvement.}

%%%%%%%%%%%%%%%%%%%%%%%%%%%%%%%%%%%%%%%%%%
\conflictsofinterest{The authors declare no conflict of interest. The founding sponsors had no role in the design of the study; in the collection, analyses, or interpretation of data; in the writing of the manuscript, and in the decision to publish the results.} 


\begin{thebibliography}{-------}
\providecommand{\natexlab}[1]{#1}

\bibitem[{Einstein}(1916)]{einstein1916}
{Einstein}, A.
\newblock {Die Grundlage der allgemeinen Relativit{\"a}tstheorie}.
\newblock {\em Ann. Phys.} {\bf 1916}, {\em 354},~769--822.

\bibitem[{Milgrom}(1983)]{milgrom1983}
{Milgrom}, M.
\newblock {A modification of the Newtonian dynamics as a possible alternative
  to the hidden mass hypothesis}.
\newblock {\em Astrophys. J.} {\bf 1983}, {\em 270},~365--370.

\bibitem[{Page}(1994)]{Page1994}
{Page}, D.N.
\newblock {Clock time and entropy.}
\newblock  In \emph{Physical Origins of Time Asymmetry}; {Halliwell}, J.J.,
  {P{\'e}rez-Mercader},~J., {Zurek}, W.H., Eds.; Cambridge University Press: Cambridge, UK,  1994; pp. 287--298.
 
\bibitem[{Verlinde}(2016)]{Verlinde2016}
\scalebox{0.96}[1]{{Verlinde}, E.P. {Emergent Gravity and the Dark Universe}. {\em SciPost Phys.} {\bf 2017}, doi:10.21468/SciPostPhys.2.3.016.}

\bibitem[{Pav{\'o}n} and {Radicella}(2014)]{Pavon2014}
{Pav{\'o}n}, D.; {Radicella}, N.
\newblock {Why We Need Dark Energy}. 
\newblock In \emph{Accelerated Cosmic Expansion}; Moreno González, C., Madriz Aguilar, J.E., Reyes Barrera, L.M., Eds.; Springer: Berlin, Germany,  2014; Astrophysics and Space Science Proceedings Volume 38, p. 143.


\bibitem[{Pandey}(2017)]{Pandey2017}
{Pandey}, B.
\newblock {Does Information Entropy Play a Role in the Expansion and
  Acceleration of the Universe?}
\newblock {\em arXiv} {\bf 2017}, arXiv:1705.08945.

\bibitem[{Hobson} \em{et~al.}(2006){Hobson}, {Efstathiou}, and
  {Lasenby}]{hobson2006}
{Hobson}, M.P.; {Efstathiou}, G.P.; {Lasenby}, A.N.
\newblock {\em {General Relativity}}; Cambridge University Press:  Cambridge, UK, 2006.

\bibitem[{Dicke}(1957)]{Dicke1957}
{Dicke}, R.H.
\newblock {Gravitation without a Principle of Equivalence}.
\newblock {\em Rev. Mod. Phys.} {\bf 1957}, {\em 29},~363--376.

\bibitem[{Chiba}(2011)]{Chiba2011}
{Chiba}, T.
\newblock {The Constancy of the Constants of Nature: Updates}.
\newblock {\em Prog. Theor. Phys.} {\bf 2011}, {\em
  126},~993--1019.

\bibitem[{Egan} and {Lineweaver}(2010)]{egan2010}
{Egan}, C.A.; {Lineweaver}, C.H.
\newblock {A Larger Estimate of the Entropy of the Universe}.
\newblock {\em Astrophys. J.} {\bf 2010}, {\em 710}, 1825--1834.

\bibitem[{Bekenstein}(1973)]{bekenstein1973}
{Bekenstein}, J.D.
\newblock {Black Holes and Entropy}.
\newblock {\em Phys. Rev. D} {\bf 1973}, {\em 7},~2333--2346.

\bibitem[{Hawking}(1976)]{hawking1976}
{Hawking}, S.W.
\newblock {Black holes and thermodynamics}.
\newblock {\em Phys. Rev. D} {\bf 1976}, {\em 13},~191--197.

%\bibitem[{Fukugita} and {Peebles}(2004)]{fukugita2004}
%{Fukugita}, M.; {Peebles}, P.J.E.
%\newblock {The Cosmic Energy Inventory}.
%\newblock {\em Astrophys. J.} {\bf 2004}, {\em 616},~643--668.

%\bibitem[{Shull} \em{et~al.}(2012){Shull}, {Smith}, and {Danforth}]{shull2012}
%{Shull}, J.M.; {Smith}, B.D.; {Danforth}, C.W.
%\newblock {The Baryon Census in a Multiphase Intergalactic Medium: 30\% of the
%  Baryons May Still be Missing}.
%\newblock {\em  Astrophys. J.} {\bf 2012}, {\em 759},~23, doi:10.1088/0004-637X/759/1/23.

%\bibitem[{Schramm} and {Turner}(1998)]{schramm1998}
%{Schramm}, D.N.; {Turner}, M.S.
%\newblock {Big-bang nucleosynthesis enters the precision era}.
%\newblock {\em Rev. Mod. Phys.} {\bf 1998}, {\em 70},~303--318.


\bibitem[{Betoule} \em{et~al.}(2014){Betoule}, {Kessler}, {Guy}, {Mosher},
  {Hardin}, {Biswas}, {Astier}, {El-Hage}, {Konig}, {Kuhlmann}, {Marriner},
  {Pain}, {Regnault}, {Balland}, {Bassett}, {Brown}, {Campbell}, {Carlberg},
  {Cellier-Holzem}, {Cinabro}, {Conley}, {D'Andrea}, {DePoy}, {Doi}, {Ellis},
  {Fabbro}, {Filippenko}, {Foley}, {Frieman}, {Fouchez}, {Galbany}, {Goobar},
  {Gupta}, {Hill}, {Hlozek}, {Hogan}, {Hook}, {Howell}, {Jha}, {Le Guillou},
  {Leloudas}, {Lidman}, {Marshall}, {M{\"o}ller}, {Mour{\~a}o}, {Neveu},
  {Nichol}, {Olmstead}, {Palanque-Delabrouille}, {Perlmutter}, {Prieto},
  {Pritchet}, {Richmond}, {Riess}, {Ruhlmann-Kleider}, {Sako}, {Schahmaneche},
  {Schneider}, {Smith}, {Sollerman}, {Sullivan}, {Walton}, and
  {Wheeler}]{betoule2014}
{Betoule}, M.; {Kessler}, R.; {Guy}, J.; {Mosher}, J.; {Hardin}, D.; {Biswas},
  R.; {Astier}, P.; {El-Hage}, P.; {Konig}, M.; {Kuhlmann}, S.; et al.
\newblock {Improved cosmological constraints from a joint analysis of the
  SDSS-II and SNLS supernova samples}.
\newblock {\em Astron.~Astrophys.} {\bf 2014}, {\em 568},~A22.

\bibitem[{Planck Collaboration} \em{et~al.}(2016){Planck Collaboration}, {Ade},
  {Aghanim}, {Arnaud}, {Ashdown}, {Aumont}, {Baccigalupi}, {Banday},
  {Barreiro}, {Bartlett}, and et~al.]{planck2015}
{Planck Collaboration}.; {Ade}, P.A.R.; {Aghanim}, N.; {Arnaud}, M.; {Ashdown},
  M.; {Aumont}, J.; {Baccigalupi},~C.; {Banday}, A.J.; {Barreiro}, R.B.;
  {Bartlett}, J.G.; et~al.~{Planck 2015 results. XIII. Cosmological parameters}.
\newblock {\em Astron.~Astrophys.} {\bf 2016}, {\em 594},~A13.




\end{thebibliography}
\end{document}